# Toward less conservative distributed stability analysis of power systems via matrix-valued differential passivity indices

Xi Ru[1], Cong Fu[2], Zhongze Li[1], Xiaoyu Peng[1], Feng Liu[1,*]

1 Department of Electrical Engineering, Tsinghua University, Beijing, 100084, China

2 Power Dispatching Control Center, Guangdong Power Grid Co., Ltd., Guangzhou 528322, China

*Corresponding author Feng Liu; lfeng@tsinghua.edu.cn

## ABSTRACT

Passivity indices have been widely adopted to derive distributed stability certificates for power systems. Nevertheless, conventional passivity indices remain scalar-valued even for multi-input-multi-output (MIMO) systems, which can introduce excessive conservatism and compromise analysis accuracy. To overcome these limitations, this paper extends the differential passivity index to a matrix-valued formulation that captures both channel-wise passivity properties and inter-channel coupling effects in MIMO subsystems. On this basis, semi-distributed and fully distributed stability criteria are developed for power systems with heterogeneous nonlinear devices. It is shown that system stability is guaranteed when the aggregate passivity excess of devices compensates for the passivity shortage imposed by the network. Furthermore, analytical passivity matrix expressions for typical power system components are derived, facilitating compositional stability analysis. Case studies on a three-bus system and a modified IEEE 118-bus system validate the effectiveness of the proposed framework.

## KEYWORDS

Power system stability, distributed stability analysis, differential passivity, passivity index, matrix-valued passivity index.

The increasing scale, heterogeneity, and dynamic complexity of modern power systems have posed significant challenges to stability analysis and dynamic performance assessment. Traditional approaches, including simulation [1], eigenanalysis [2], and energy-function methods [3], have been well developed and widely applied in conventional synchronous-generator-dominated systems, where the dynamics are relatively homogeneous and centrally manageable [4]. However, modern power systems are undergoing a profound transformation driven by the rapid integration of renewable generation, power electronic devices, and distributed energy resources [5], [6]. This transition introduces diverse and fast-varying dynamics, substantially increases the number of dynamic buses, and renders centralized stability assessment increasingly impractical owing to its heavy computational burden, limited scalability, and difficulty in localizing instability sources [7], [8]. These challenges motivate the development of distributed stability analysis methods for large-scale heterogeneous power systems.

Among various methods for analyzing power system stability, passivity and dissipativity theories have garnered considerable attention in recent years. Passivity is typically defined as an energy-based property in which the energy absorbed from the environment is greater than or equal to the energy stored in the system [9]. It provides a general framework for assessing system stability: a system has a stable equilibrium if it is passive and admits a positive definite storage function [10]. Moreover, a system is passive if all of its interconnected subsystems, through feedback, are passive [11]. This compositional nature enables a distributed stability assessment, effectively avoiding the redundancy inherent in centralized analysis [12]–[14]. To quantify the passivity surplus or shortage of a subsystem, passivity indices have been introduced and applied in stability analysis and controller design [15]–[17]. For linear time-invariant systems, their relation to positive realness further facilitates small-signal stability verification [18]. Owing to these advantages, passivity-based analysis has attracted increasing attention in power systems, especially in inverter-dominated scenarios [19]–[21].

Despite these advantages, existing passivity-based methods still face important limitations when applied to modern heterogeneous power systems. A critical limitation is that most existing approaches rely on scalar passivity indices [14], [22], [23]. While effective for single-input single-output systems or weakly coupled settings, scalar indices are inherently limited in characterizing the full passivity properties of multi-input multi-output (MIMO) subsystems. In particular, they cannot explicitly capture the coupling and complementarity among different input-output channels, which may lead to conservative stability conditions and may even fail to provide a valid characterization.

To address these shortcomings, recent research has extended scalar-valued passivity indices to matrix-valued formulations based on the classical concept of passivity [24]. Building upon this foundation, this paper further extends the concept to **differential passivity** and develops a comprehensive stability analysis framework based on matrix-valued differential passivity indices. The main contributions of this paper are summarized as follows:

1) Theoretical Extension: We extend the classical scalar differential passivity index to a matrix-valued form for MIMO subsystems in power systems. Compared with scalar indices, the proposed matrix-valued indices capture not only channel-wise passivity properties but also the coupling and complementarity among different input-output channels, thereby providing a more expressive characterization of subsystem passivity in heterogeneous interconnected power systems.

2) Improved Stability Criteria: Based on the proposed matrix-valued differential passivity indices, we derive distributed stability criteria for interconnected power systems. In particular, the semi-distributed criterion achieves improved accuracy and reduced conservativeness compared with conventional scalar-index-based criteria, while the fully distributed criterion provides a simpler distributed alternative. We further develop an application framework for power systems, including the derivation of differential passivity matrices for representative device and network models and a two-level distributed stability assessment procedure.

The rest of this paper is organized as follows. Section 1 presents the modeling of the interconnected power system. Section 2 introduces the matrix-valued differential passivity indices, and then Section 3 derives the distributed stability criteria. Section 4 applies the proposed theory to power-system models. Section 5 presents case studies, and Section 6 concludes the paper.

*Notations*: For any real matrix $A$, its symmetric part is denoted by $\bar{A} := (A + A^T)/2$.

## 1. Power systems modeling

### 1.1 Bus dynamics model

Consider a network-reduced power system with $N$ buses and transmission lines. The power system can be abstracted as an undirected graph $\mathcal{G} = (\mathcal{V}, \mathcal{E})$, where $\mathcal{V}$ is the set of buses and $\mathcal{E}$ is the set of transmission lines. For each bus $i \in \mathcal{V}$, $V_i$, $\theta_i$, $P_i$, and $Q_i$ denote its voltage amplitude, phase angle, active and reactive power injection, respectively. The bus dynamics is modeled in a general form as below:

$$\begin{cases} \dot{x}_i = f_i\left(x_i, u_{\mathrm{dev},i}\right) \\ y_{\mathrm{dev},i} = C_i x_i \end{cases} \tag{1}$$

where $x_i \in \mathbb{R}^{n_i}$ is the state variable that contains at least $\theta_i$ and $V_i$. The input and output variables are selected as $u_{\mathrm{dev},i} := -[P_i, Q_i / V_i]^T$ and $y_{\mathrm{dev},i} = C_i x_i := [\theta_i, V_i]^T$, where the subscript dev indicates the device dynamics. The map $f_i : \mathcal{D}_i \times \mathbb{R}^2 \to \mathbb{R}^{n_i}$ is locally Lipschitz, where $\mathcal{D}_i$ is a domain that contains the equilibrium point of Eq. (1).

Many common dynamic devices in power systems conform to the structure of Eq. (1). First, we consider the third-order flux-decay model of SGs [25]:

$$\begin{cases} \dot{\theta}_i = \omega_i \\ M_i \dot{\omega}_i = -D_i \omega_i - P_i + P_i^g \\ T_{di'} \dot{E}_{qi'} = -E_{qi'} - (x_{di} - x_{di'}) \dfrac{Q_i}{E_{qi'}} + E_{fi} \end{cases} \tag{2}$$

Here, $E_{qi} \angle \theta_i$ denotes the complex $q$-axis transient internal voltage, and $\omega_i$ denotes the rotor frequency deviation. The constants $M_i > 0$, $D_i > 0$, and $T_{di'} > 0$ represent the moment of inertia, damping coefficient, and the $d$-axis open-circuit transient time constant, respectively. $x_{di}$, $x_{di'}$ represents the $d$-axis synchronous and transient reactance, respectively. $P_i^g$ and $E_{fi}$ are control signals.

Since the integration of renewable generation is primarily interfaced through inverters, we next introduce the dynamic model of conventional droop-controlled inverters (CDs) with additional cross-loop controls $u_{\theta V,i}$ and $u_{V\theta,i}$ as below [26].

$$\begin{cases} \tau_{1i} \dot{\theta}_i = -(\theta_i - \theta_i^*) - D_{1i}(P_i - P_i^*) - u_{\theta V,i} \\ \tau_{2i} \dot{V}_i = -(V_i - V_i^*) - D_{2i}(Q_i - Q_i^*) - u_{V\theta,i} \end{cases} \tag{3}$$

where $\tau_{1i}$ and $\tau_{2i}$ are the time constants; $D_{1i}$ and $D_{2i}$ are the droop gains; $u_{\theta V,i}$ and $u_{V\theta,i}$ are possible control signals.

Another type of inverter is the quadratic droop-controlled inverter [27]. Its dynamic behavior is described by the following model.

$$\begin{cases}\tau_{1i}\dot{\theta}_i = -(\theta_i - \theta_i^*) - D_{1i}(P_i - P_i^*) - u_{\theta V,i} \\ \tau_{2i}\dot{V}_i = -V_i(V_i - u_i^*) - D_{2i}Q_i - u_{V\theta,i}\end{cases} \tag{4}$$

where the control input $u_i^* = D_{2i}Q_i^* / V_i^* + V_i^*$ is a constant determined by the steady-state conditions. $u_{\theta V,i}$ and $u_{V\theta,i}$ are control signals.

## 1.2 Power network model

The bus dynamics are interconnected through transmission lines that can be modeled using power flow equations. The network topology is represented by the admittance matrix $Y = G + jB$. For each bus $i \in \mathcal{V}$, the power flow equations are:

$$\begin{cases}P_i = V_i \sum_{j\in\mathcal{V}} V_j (G_{ij}\cos\theta_{ij} + B_{ij}\sin\theta_{ij}) \\ Q_i = V_i \sum_{j\in\mathcal{V}} V_j (G_{ij}\sin\theta_{ij} - B_{ij}\cos\theta_{ij})\end{cases} \tag{5}$$

where $G_{ij}$ and $B_{ij}$ are the components of matrix $Y$, $\theta_{ij} := \theta_i - \theta_j$ is the phase difference between bus $i$ and bus $j$.

We denote the set of input-output pairs of all devices by $(u_{\mathrm{dev}}, y_{\mathrm{dev}})$, and the input-output pair of the network by $(u_{\mathrm{net}}, y_{\mathrm{net}})$, where the subscript $\mathrm{net}$ indicates the network model. Their relationship is defined as follows.

$$\begin{cases}y_{\mathrm{net}} = -u_{\mathrm{dev}} := (P_1, Q_1/V_1, \cdots, P_n, Q_n/V_n)^T \\ u_{\mathrm{net}} = y_{\mathrm{dev}} := (\theta_1, V_1, \cdots, \theta_n, V_n)^T\end{cases} \tag{6}$$

The power flow function (5) can be rewritten as:

$$y_{\mathrm{net}} = g(u_{\mathrm{net}}) \tag{7}$$

## 1.3 Feedback-interconnection model of the overall system

Combining the device dynamics and the network power flow equation yields an overall interconnected power system model:

$$\begin{cases}\dot{x} = f(x, u_{\mathrm{dev}}) \\ y_{\mathrm{dev}} = Cx \\ y_{\mathrm{net}} = g(u_{\mathrm{net}})\end{cases} \tag{8}$$

where $x := [x_1, x_2, \cdots, x_N]^T$. Its input-output feedback-interconnection configuration is illustrated in Fig. 1.

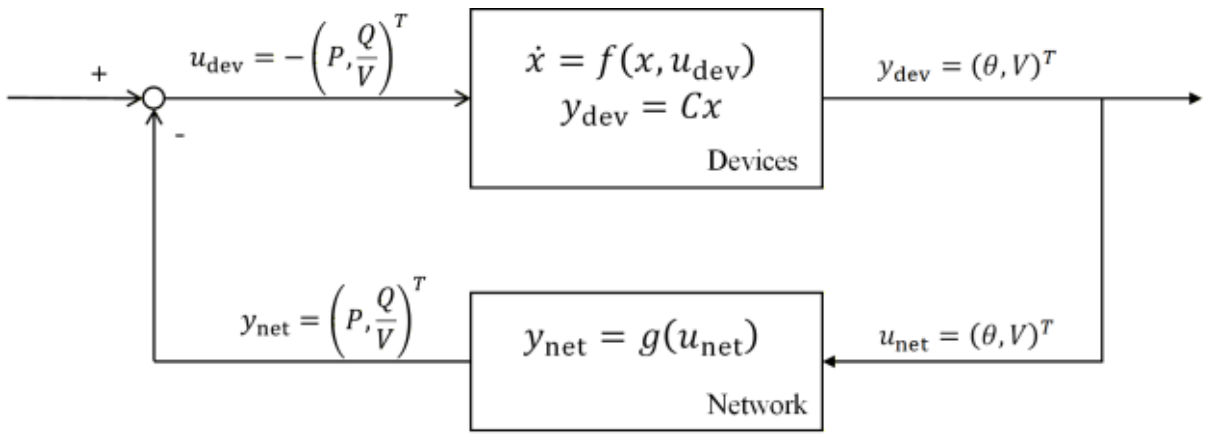

**Fig. 1 Feedback-interconnection model of power systems.**

**Definition 1**. $(x^*, u_{\mathrm{dev}}^*, y_{\mathrm{dev}}^*)$ is called an equilibrium triplet of system (8) if

$$\begin{cases}0 = f(x^*, u_{\mathrm{dev}}^*) \\ y_{\mathrm{dev}}^* = Cx^* \\ u_{\mathrm{dev}}^* = -g(y_{\mathrm{dev}}^*)\end{cases} \tag{9}$$

where $x^*$ is the equilibrium point, $u^*_{\text{dev}}$ and $y^*_{\text{dev}}$ are the corresponding input and output at $x^*$.

Throughout the paper, we assume the following general assumption:

**Assumption 1**. The equilibrium $x^*$ exists and is isolated.

In the following parts, all analysis is carried out in equilibrium-shifted coordinates with respect to the equilibrium triplet $(x^*, u^*_{\text{dev}}, y^*_{\text{dev}})$. Specifically, define the shifted variables as $\hat{x} := x - x^*$, $\hat{u}_{\text{dev}} := u_{\text{dev}} - u^*_{\text{dev}}$, $\hat{y}_{\text{dev}} := y_{\text{dev}} - y^*_{\text{dev}}$. The network-side variables $u_{\text{net}}$ and $y_{\text{net}}$ are defined consistently in the shifted coordinates through the interconnection relation (6). For notational simplicity, in the remainder of the paper, these shifted variables are still denoted by $x$, $u_{\text{dev}}$, $u_{\text{net}}$, $y_{\text{dev}}$ and $y_{\text{net}}$, unless otherwise specified.

## 2 From conventional passivity index to differential passivity matrix

### 2.1 Classical concepts of passivity indices and their limitations

Passivity theory provides an energy-based framework for analyzing the stability of interconnected dynamical systems, and has been widely applied to power system analysis. Consider a nonlinear system with input $u \in \mathbb{R}^m$, output $y \in \mathbb{R}^m$ and state $x \in \mathbb{R}^n$. The system is said to be passive if there exists a continuously differentiable storage function $V(x) \geq 0$ such that

$$u^T y \geq \dot{V} = \frac{\partial V}{\partial x} f(x,u), \forall (x,u) \in \mathbb{R}^n \times \mathbb{R}^m \tag{10}$$

Moreover, it is input-feedforward passive if $u^T y \geq \dot{V} + u^T \varphi(u)$ for some function $\varphi$, and output-feedback passive if $u^T y \geq \dot{V} + y^T \rho(y)$ for some function $\rho$. It is strictly passive if $u^T y \geq \dot{V} + \psi(x)$ for some positive definite (p.d.) function $\psi$. If $\varphi(u) = \varepsilon u$ and $\rho(y) = \gamma y$ with $\varepsilon, \gamma \in \mathbb{R}$, the scalar numbers $\varepsilon$ and $\gamma$ are called the input-feedforward and the output-feedback passivity index, respectively [15]. A positive $\varepsilon$ or $\gamma$ indicates passivity excess in the system, while a negative value indicates passivity shortage.

However, scalar passivity indices are inherently restrictive for multi-input multi-output (MIMO) systems. Since they do not capture inter-channel coupling and complementarity across devices, the stability assessments can be can be overly conservative. This motivates us to seek a more expressive formulation.

### 2.2 Differential passivity and its extension to matrix-valued indices

In the literature, it has been recognized that differential passivity offers advantages over conventional passivity in power system stability analysis. Here, we focus on two specific types of differential passivity: output-differential passivity (ODP), which applies to bus dynamics, and input-differential passivity (IDP), which applies to networks. Compared to conventional concepts of ODP and IDP [14], we use matrix-valued indices instead of scalar ones here.

**Definition 2** (ODP of bus dynamics). The system (1) is said to be **output-differential passive (ODP)** if there exists a matrix $\mathcal{S}_{\text{dev},i} \in \mathbb{R}^{m \times m}$ and a continuously differentiable positive semi-definite (p.s.d.) storage function $V_i(x)$ such that

$$\dot{V}_i \leq u^T_{\text{dev},i} \dot{y}_{\text{dev},i} - y^T_{\text{dev},i} \mathcal{S}^T_{\text{dev},i} \dot{y}_{\text{dev},i} \tag{11}$$

Moreover, it is strictly output-differential passive (SODP) if $\dot{V}_i \leq u^T_{\text{dev},i} \dot{y}_{\text{dev},i} - y^T_{\text{dev},i} \mathcal{S}^T_{\text{dev},i} \dot{y}_{\text{dev},i} - \psi_i(\dot{y}_{\text{dev},i})$ for some p.d. function $\psi_i$.

In Definition 2, the matrix $\mathcal{S}_{\text{dev},i}$ is a **matrix-valued output-differential passivity index**, which is referred to as an **output-differential passivity matrix (ODPM)** of the system.

The ODP of the system $\mathcal{G}_{\text{dev},i} : u_{\text{dev},i} \mapsto y_{\text{dev},i}$ is equivalent to the passivity of the system

$\tilde{\mathcal{G}}_{\mathrm{dev},i} : \tilde{u}_{\mathrm{dev},i} \to \tilde{y}_{\mathrm{dev},i}$, where $\tilde{u}_{\mathrm{dev},i} = u_{\mathrm{dev},i} - \mathcal{S}_{\mathrm{dev},i} y_{\mathrm{dev},i}$ and $\tilde{y}_{\mathrm{dev},i} = \dot{y}_{\mathrm{dev},i}$, as shown in Fig. 2. We refer to $\tilde{\mathcal{G}}_{\mathrm{dev},i}$ as the output-differential equivalent system of $\mathcal{G}_{\mathrm{dev},i}$.

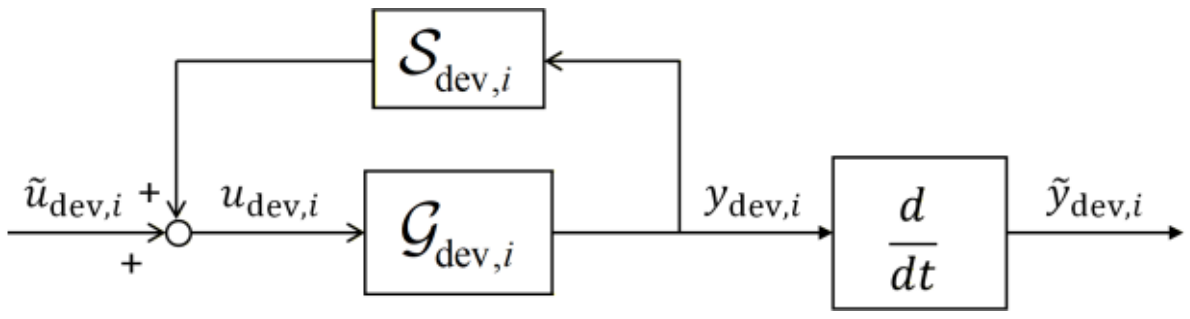

**Fig. 2 The block diagram of output-differential passivity.**

The input-differential passivity is the dual concept of the output-differential passivity.

**Definition 3** (IDP for power networks). The system (7) is said to be input-differential passive if there exists a matrix $\mathcal{S}_{\mathrm{net}} \in \mathbb{R}^{m\times m}$ such that $\dot{u}_{\mathrm{net}}^T y_{\mathrm{net}} - \dot{u}_{\mathrm{net}}^T \mathcal{S}_{\mathrm{net}} u_{\mathrm{net}} \geq 0$.

In Definition 3, the matrix $\mathcal{S}_{\mathrm{net}}$ is a **matrix-valued input differential passivity index**, which is referred to as an **input-differential passivity matrix (IDPM)** of the system.

The input-differential passivity of the system $\mathcal{G}_{\mathrm{net}} : u_{\mathrm{net}} \mapsto y_{\mathrm{net}}$ is equivalent to the passivity of the system $\tilde{\mathcal{G}}_{\mathrm{net}} : \tilde{u}_{\mathrm{net}} \to \tilde{y}_{\mathrm{net}}$ with the input $\tilde{u} = \dot{u}$ and the output $\tilde{y}_{\mathrm{net}} = y_{\mathrm{net}} - \mathcal{S}_{\mathrm{net}} u_{\mathrm{net}}$ as shown in Fig. 3. $\tilde{\mathcal{G}}_{\mathrm{net}}$ is called the input-differential equivalent system of $\mathcal{G}_{\mathrm{net}}$.

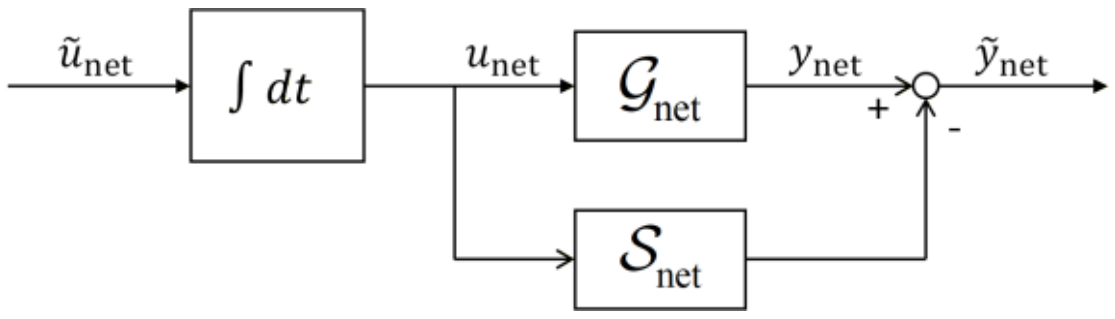

**Fig. 3 The block diagram of input-differential passivity.**

The differential passivity matrix $\mathcal{S}_{\mathrm{dev},i}$ or $\mathcal{S}_{\mathrm{net}}$ can be interpreted as an equivalent virtual output feedback loop or input feedforward loop applied to the system, through which the excess or shortage of passivity inherent in the system is removed. Hence, $\mathcal{S}_{\mathrm{dev},i}$ or $\mathcal{S}_{\mathrm{net}}$ serves as a matrix-valued passivity index whose elements reflect the passivity measure (either excess or shortage) of corresponding input–output channels of the system. When $\mathcal{S}_{\mathrm{dev},i}$ or $\mathcal{S}_{\mathrm{net}}$ is a diagonal matrix with identical diagonal elements, it is directly reduced to the classical scalar passivity index in [14].

## 3 Distributed stability criteria empowered by differential passivity matrices

### 3.1 Differential passivity matrix–based stability conditions

We first present several passivity conditions that subsequent analyses may require.

**Condition 1** (ODP for bus dynamics). Each bus $i \in \mathcal{V}$ with dynamic (1) satisfies the ODP property with a symmetric ODPM $\mathcal{S}_{\mathrm{dev},i} \in \mathbb{R}^{2\times 2}$ and a p.d. storage function $V_i$.

**Condition 2** (SODP for bus dynamics). Each bus $i \in \mathcal{V}$ with dynamic (1) satisfies the SODP property with a symmetric ODPM $\mathcal{S}_{\mathrm{dev},i} \in \mathbb{R}^{2\times 2}$ and a p.d. storage function $V_i$.

**Condition 3** (IDP for networks). The network (7) satisfies the IDP property with a symmetric IDPM $\mathcal{S}_{\mathrm{net}} \in \mathbb{R}^{2N\times 2N}$.

Since Condition 1 is implied by Condition 2, the following equivalent system construction assumes only the former. When Condition **1** is satisfied for power system (8), the ODPM $\mathcal{S}_{\mathrm{dev}}$ is composed of the ODPMs of buses $1 \sim N$:

$$\mathcal{S}_{\mathrm{dev}} := \mathrm{diag}(\mathcal{S}_{\mathrm{dev},1}, \ldots, \mathcal{S}_{\mathrm{dev},N}) \tag{12}$$

Furthermore, the power system (8) satisfying Conditions 1 and 3 can be reformulated using the output-differential equivalent systems of bus dynamics and the input-differential equivalent system of the network, as shown in Fig. 4. In this way, the extra feedback subsystem $\mathcal{G}_{\text{ext}}$ ensures the input–output consistency of feedback interconnection.

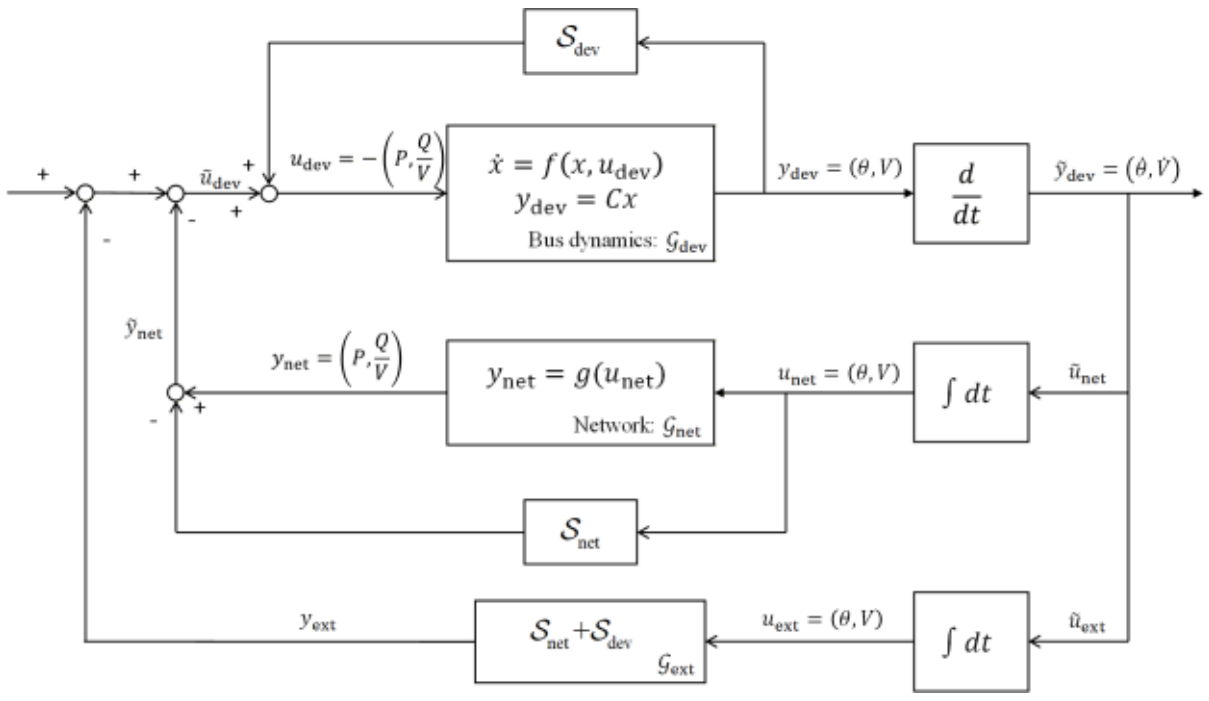

**Fig. 4 Differential equivalent interconnection model of the power system.**

### 3.2 Distributed stability criteria

Since the differential equivalent system shares the same zero-input dynamics as the original system (Fig. 1), analyzing its passivity provides direct insight into the original system's zero-input stability, as formalized in the following theorems.

**Theorem 1**. Suppose that Assumption 1 holds for power system (8). If Conditions 1 and 3 are satisfied with

$$\mathcal{S}_{\text{net}} + \mathcal{S}_{\text{dev}} \succ 0 \tag{13}$$

Then the equilibrium $x = 0$ of the shifted system is Lyapunov stable. Furthermore, if Condition 2 is satisfied, and for each bus dynamics (1), a constant output implies a constant state, i.e. $\dot{y}_{\text{dev},i} = 0, \forall t \geq 0 \Rightarrow \dot{x}_i = 0, \forall t \geq 0$, then the equilibrium $x = 0$ is asymptotically stable.

Proof: Let $V_{\text{dev}} = \sum_i V_i(x_i)$ and $V_{\text{net}} = \frac{1}{2} u_{\text{net}}^T (\mathcal{S}_{\text{net}} + \mathcal{S}_{\text{dev}}) u_{\text{net}}$. Consider the Lyapunov candidate:

$$V(x) = V_{\text{dev}} + V_{\text{net}} = \sum_i V_i(x_i) + \frac{1}{2} y_{\text{dev}}^T (\mathcal{S}_{\text{net}} + \mathcal{S}_{\text{dev}}) y_{\text{dev}} \tag{14}$$

Since $V_i$ is positive definite, $V_{\text{dev}} > 0$, $\forall x \neq 0$. In addition, $\mathcal{S}_{\text{net}} + \mathcal{S}_{\text{dev}} \succ 0$ denotes that $V_{\text{net}} \geq 0$. Hence, there exists a $\delta$-neighborhood $D := \left\{ x \in \mathbb{R}^n : \|x\| < \delta \right\}$ of $x = 0$ such that $V(0) = 0$ and $V(x) > 0$, $\forall x \in D \backslash \{0\}$. That is, $V(x)$ is a positive definite function in $D$. According to condition 3 and the relation $u_{\text{net}} = y_{\text{dev}}$, $y_{\text{net}} = -u_{\text{dev}}$, we have $\dot{y}_{\text{dev}}^T u_{\text{dev}} \leq -\dot{y}_{\text{dev}}^T \mathcal{S}_{\text{net}} y_{\text{dev}}$. Denote that $\mathcal{S} = \mathcal{S}_{\text{dev}} + \mathcal{S}_{\text{net}}$. Since $\mathcal{S}$ is symmetrical, the time derivative of $V$ is calculated as:

$$\begin{aligned}
\dot{V} &= \sum_i \dot{V}_i + y_{\text{dev}}^T \mathcal{S} \dot{y}_{\text{dev}} \\
&\leq \sum_i \left( u_{\text{dev},i}^T \dot{y}_{\text{dev},i} - y_{\text{dev},i}^T \mathcal{S}_{\text{dev},i}^T \dot{y}_{\text{dev},i} \right) + y_{\text{dev}}^T \mathcal{S} \dot{y}_{\text{dev}} \\
&= u_{\text{dev}}^T \dot{y}_{\text{dev}} + \dot{y}_{\text{dev}}^T \left( \mathcal{S} - \mathcal{S}_{dev} \right) y_{\text{dev}} \\
&= u_{\text{dev}}^T \dot{y}_{\text{dev}} + \dot{y}_{\text{dev}}^T \mathcal{S}_{\text{net}} y_{\text{dev}} = 0
\end{aligned}$$

Thus, $V(x)$ is a Lyapunov function in $D$, which implies $x = 0$ is stable. Furthermore, if Condition 2 holds, we have $\dot{V} \leq -\sum_i \psi_i(\dot{y}_{\text{dev},i}) \leq 0$. By Assumption 1, $x = 0$ is an isolated equilibrium. Hence, there exists a $\mu > 0$ such that the set $\Xi := \left\{ x \in D : V(x) \leq \mu \right\}$ is positive invariant and contains no other equilibrium point except $x = 0$. Let $\Gamma = \Xi \cap \left\{ x \in D : \dot{V}(x) = 0 \right\}$. Since $\psi_i$ is positive definite, $\dot{V}(x) = 0$ implies $\dot{y}_{\text{dev}} = 0$. By the assumed implication $\dot{y}_i = 0 \Rightarrow \dot{x}_i = 0$, it follows that any trajectory contained in $\Gamma$ must satisfy $\dot{x}(t)=0$. Thus, no solution can stay identically in $\Gamma$ other than the equilibrium point $x=0$. According to the LaSalle's invariance principle, $x=0$ is asymptotically stable. □

Theorem 1 offers a semi-distributed stability criterion by combining distributed computation and centralized verification. Specifically, individual devices and the network can independently verify their own passivity conditions (Conditions 1 and 3) using local models and measurements only. System-wide stability, however, still relies on a centralized check of the matrix condition $\mathcal{S}_{\text{net}}+\mathcal{S}_{\text{dev}}\succ 0$. Theorem 1 follows the standard passivity principle: the feedback interconnection of passive systems remains passive. Since both the bus dynamics and the network are passive under the differential equivalent formulation, the overall system is passive if the extra feedback path is passive, which further implies zero-input stability, as the p.d. storage function naturally serves as a Lyapunov function.

To further decentralize the stability criterion, we define scalar passivity indices based on the minimum eigenvalues of the symmetric DPMs. Specifically, $\sigma(\mathcal{S}_{\text{dev},i}):=\lambda_{\min}(\mathcal{S}_{\text{dev},i})$ and $\sigma(\mathcal{S}_{\text{net}}):=\lambda_{\min}(\mathcal{S}_{\text{net}})$ are introduced to quantify the passivity excess or shortage of bus dynamics and the network, respectively. These quantitative passivity indices allow individual subsystems to independently assess their contribution, leading to the following fully distributed stability criterion.

**Theorem 2**. Let Assumption 1 hold for the power system (8). Assume that Conditions 1 and 3 are satisfied. If the following inequality holds:

$$\min_i[\sigma(\mathcal{S}_{\text{dev},i})]+\sigma(\mathcal{S}_{\text{net}})>0, \tag{15}$$

Then the equilibrium $x=0$ of the shifted system is Lyapunov stable. Furthermore, if the system satisfies Condition 2, and for each bus dynamics (1), a steady output implies a steady state, i.e. $\dot{y}_i=0,\forall t\geq 0\Rightarrow \dot{x}_i=0,\forall t\geq 0$, then $x=0$ is asymptotically stable.

The result follows directly from Theorem 1 and the eigenvalue bound of symmetric matrices $\lambda_{\min}(A+B)\geq\lambda_{\min}(A)+\lambda_{\min}(B)$.

Theorem 2 decomposes the system-level stability condition (13) into device-level conditions. Each subsystem $i$ only needs to verify whether $\sigma(\mathcal{S}_{\text{dev},i})+\sigma(\mathcal{S}_{\text{net}})>0$ when assessing system-wide stability. The minimum eigenvalue $\sigma(\mathcal{S}_{\text{dev},i})$ can be regarded as a scalar index characterizing the local passivity surplus or shortage of the subsystem. In this sense, the stability criterion based on scalar differential passivity indices [14] is a special case of Theorem 2, applicable when the differential passivity matrices of devices are fully diagonal. Although the fully distributed stability criterion simplifies the stability assessment process, it can introduce additional conservativeness. In practice, a trade-off can be made between accuracy and efficiency by selecting an appropriate form of the stability criteria.

## 4 Application to power systems

To demonstrate the applicability of the proposed stability criteria, we first show that the required symmetric differential passivity matrices can be rigorously obtained in a practically relevant special case, namely, without device-side cross control and under lossless networks. We then extend the framework to more general engineering settings, where the resulting differential passivity matrices are generally nonsymmetric, and their symmetric parts are used for approximate stability assessment. Based on the above analysis, a systematic procedure is further proposed for evaluating the stability of general power systems using the developed passivity-based criteria.

### 4.1 ODP verification for individual bus dynamics

In this subsection, we show that the three typical dynamic devices introduced in Section 1.1 can satisfy the ODP property through appropriate control design. The following results can be verified by direct substitution into Definition 2, and the proofs are therefore omitted for brevity.

1) Synchronous Generators (SGs): Consider the SG model (2). We first consider the control signals with standard PID-like self-loop terms only, that is:

$$\begin{aligned} P_i^g &= P_{gi}^{self} = -K_{Ii}\int_0^\tau \omega_i \mathrm{d}t - K_{Pi}\omega_i + P_i^{g*} \\ E_{fi} &= E_{fi}^{self} = -K_{Ei}\left(E_{qi}^{'} - E_{qi}^{'*}\right) + E_{fi}^{*} \end{aligned} \tag{16}$$

where $K_{Ii}$, $K_{Pi}$, $K_{E,i}$ are control parameters of the self-loop control, $P_{gi}^{*}$ and $E_{fi}^{*}$ are steady-state inputs. The resulting ODPM takes the following diagonal form:

$$\mathcal{S}_{\mathrm{dev},i} = \begin{bmatrix} \sigma_1 & \\ & \sigma_2 \end{bmatrix} \tag{17}$$

where $\sigma_1 < K_{Ii}, \sigma_2 < (K_{Ei}+1)/(x_{di} - x_{di'})$. Under the above parameter conditions, the ODPM in (17) is diagonal and hence symmetric. Therefore, Conditions 1 and 2 are rigorously satisfied for the SG model in this no-cross-control case.

We next consider the case with cross-loop control [28], [29], which is introduced here to compensate for the network coupling deficiency and thereby enhance the overall system stability [30].

$$\begin{aligned} P_{gi}^{cross} &= -K_{wV,i}(E_{qi}^{'} - E_{qi}^{'*}) \\ E_{fi}^{cross} &= -K_{V\theta,i}\int \omega_i \mathrm{d}t - K_{V\omega,i}\omega_i \end{aligned} \tag{18}$$

where $K_{\omega V,i}$, $K_{V\theta,i}$ and $K_{V\omega,i}$ are control parameters of cross-loop control. The complete control signals are then given by

$$P_{gi} = P_{gi}^{self} + P_{gi}^{cross}, \quad E_{fi} = E_{fi}^{self} + E_{fi}^{cross} \tag{19}$$

Then, the ODPM is analytically expressed in Proposition 1.

**Proposition 1**. Let $\sigma_1 < K_{Ii}, \sigma_2 < (K_{Ei}+1)/(x_{di} - x_{di'})$. The SG model (2) with the control law (19) is SODP at the equilibrium $x^*$ with the following ODPM:

$$\mathcal{S}_{\mathrm{dev},i} = \begin{bmatrix} \sigma_1 & K_{\omega V,i} \\ \dfrac{K_{V\theta,i}}{x_d - x_{d'}} & \sigma_2 \end{bmatrix} \tag{20}$$

provided that $K_{Pi} + D_i > 0,$, $x_{di} - x_{di'} > 0$ and $K_{V\omega,i} < 2\sqrt{T_{d'}(D+K_P)(x_d - x_{d'})}$.

2) Conventional Droop-Controlled Inverters (CDs): Consider the CD model (3). We first consider the control signals without cross-loop terms, i.e., $u_{\theta V,i}=0$ and $u_{V\theta,i}=0$. The resulting ODPM takes the following diagonal form:

$$\mathcal{S}_{\mathrm{dev},i} = \begin{bmatrix} \dfrac{1}{D_{1i}} & \\ & \dfrac{D_{2i}Q_i^* + V_i^*}{D_{2i}V_i^{*2}} \end{bmatrix} \tag{21}$$

when $D_{1i}, D_{2i}, \tau_{1i}, \tau_{2i} > 0$. Under the above parameter conditions, the ODPM in (21) is diagonal and hence symmetric. Therefore, Conditions 1 and 2 are rigorously satisfied for the CD model.

We next consider cross-loop control signals:

$$\begin{aligned} u_{\theta V,i} &= K_{\omega V,i}(V_i - V_i^*) \\ u_{V\theta,i} &= K_{V\theta,i}(\theta_i - \theta_i^*) + K_{V\omega,i}\dot{\theta}_i \end{aligned} \tag{22}$$

where $K_{\omega V,i}$, $K_{V\theta,i}$ and $K_{V\omega,i}$ are control parameters of cross-loop control. Then, the ODPM is given in Proposition 2 below.

**Proposition 2**. The CD inverter model (3) with the cross-loop control (22) is SODP at the equilibrium $x^*$ with the following ODPM:

$$\mathcal{S}_{\text{dev},i} = \begin{bmatrix} \frac{1}{D_{1i}} & \frac{K_{\omega V,i}}{D_{1i}} \\ \frac{K_{V\theta,i}}{D_{2i}V_i^*} & \frac{D_{2i}Q_i^* + V_i^*}{D_{2i}V_i^{*2}} \end{bmatrix} \tag{23}$$

provided that: $D_{1i}, D_{2i}, \tau_{1i}, \tau_{2i} > 0$ , $K_{V\omega,i} \le 2\sqrt{\tau_{1i}\tau_{2i}V_i^* D_{i2} / D_{1i}}$ .

3) Quadratic Droop-Controlled Inverters (QDs): Consider the quadratic droop-controlled inverters (4). We first consider the control signals without cross-loop term, that is $u_{\theta V,i}=0$ and $u_{V\theta,i}=0$ . The resulting ODPM takes the following diagonal form:

$$\mathcal{S}_{\text{dev},i} = \begin{bmatrix} \frac{1}{D_{1i}} & \\ & \frac{1}{D_{2i}} \end{bmatrix} \tag{24}$$

when $D_{1i}, D_{2i}, \tau_{1i}, \tau_{2i} > 0$ . Under the above parameter conditions, the ODPM in (24) is diagonal and hence symmetric. Therefore, Conditions 1 and 2 are rigorously satisfied for the QD model.

We next consider the same cross-loop control signals as in (22). Then, the ODPM is analytically expressed in Proposition 3.

**Proposition 3**. The QD inverter model (4) with the cross-loop control (22) is SODP at the equilibrium $x^*$ with the following ODPM:

$$\mathcal{S}_{\text{dev},i} = \begin{bmatrix} \frac{1}{D_{1i}} & \frac{K_{\omega V,i}}{D_{1i}} \\ \frac{K_{V\theta,i}}{D_{2i}V_i^*} & \frac{1}{D_{2i}} \end{bmatrix} \tag{25}$$

provided that: $D_{1i}, D_{2i}, \tau_{1i}, \tau_{2i} > 0$ , $K_{V\omega,i} \le 2\sqrt{\tau_{1i}\tau_{2i}V_i^* D_{i2} / D_{1i}}$ .

For all the above device models, the no-cross-control case yields diagonal and symmetric ODPMs (17), (21) and (24), under which Conditions 1 and 2 are rigorously satisfied. When cross-loop control is introduced, off-diagonal terms appear in the corresponding ODPMs (20), (23) and (25), capturing the coupling between different input-output channels. As a result, the ODPMs are generally no longer exactly symmetric, although they provide a richer characterization of the MIMO passivity properties of the devices. For the subsequent approximate stability assessment, the symmetric parts of these nonsymmetric ODPMs will be used in the general case.

### 4.2 IDP verification for power network coupling

Consider the LTI form of the power network model (7):

$$y_{\text{net}} = G_{\text{net}} u_{\text{net}} \tag{26}$$

where the gain matrix $G_{\text{net}} \in \mathbb{R}^{2N\times 2N}$ is a constant matrix. According to Definition 3, the linearized network-side transfer matrix $G_{\text{net}}$ provides an exact IDPM for the power network model.

To derive the exact expression of $G_{\text{net}}$ , we first consider the lossless network case. Specifically, for the power flow model (5) with $G_{ij} = 0$ in admittance matrix $Y = G + jB$ , the linearized transfer function matrix $G_{\text{net}}^{\text{lossless}}$ can be derived as follows:

$$\begin{bmatrix} P \\ Q/V \end{bmatrix} = G_{\text{net}}^{\text{lossless}} \begin{bmatrix} \theta \\ V \end{bmatrix} = \begin{bmatrix} A & D \\ D^T & C \end{bmatrix} \begin{bmatrix} \theta \\ V \end{bmatrix} \tag{27}$$

In this case, $G_{\text{net}}^{\text{lossless}}$ is symmetric and therefore can be directly used in the stability criteria. Then we consider the more general lossy network case with $G_{ij} \neq 0$. The transfer function matrix $G_{\text{net}}^{\text{lossy}}$ can be derived via linearization:

$$\begin{bmatrix} P \\ Q/V \end{bmatrix} = G_{\text{net}}^{\text{lossy}} \begin{bmatrix} \theta \\ V \end{bmatrix} = \begin{bmatrix} A' & D' \\ E' & C' \end{bmatrix} \begin{bmatrix} \theta \\ V \end{bmatrix} \tag{28}$$

The Jacobian matrix blocks in Eq. (27) and Eq. (28) are given in the Appendix. In general, $G_{\text{net}}^{\text{lossy}}$ is nonsymmetric. Nevertheless, it still serves as an exact IDPM under Definition 3. For the subsequent application of Theorems 1 and 2, we use its symmetric part as an approximate matrix in the stability assessment.

It should be noted that the input and output ordering of the linear model in Eq. (27) and Eq. (28) differs from that in Eq. (6). This difference can be remedied by a permutation transformation, as illustrated by the permutation matrix $N = [N_{ij}] \in \mathbb{R}^{2N \times 2N}$, defined as follows:

$$N_{ij} = \begin{cases} 1, & i = 2j-1, \quad j = 1, \dots, N, \\ 1, & i = 2j-2N, \quad j = N+1, \dots, 2N, \\ 0, & \text{otherwise.} \end{cases} \tag{29}$$

The transfer function matrix $G_{\text{net}}^{'}$ under the input and output selection (6) is given by $G_{\text{net}}^{'} = NG_{\text{net}}N^T$. Since permutation transformations preserve the eigenvalues and symmetry of matrices, it follows that the power network model (5) is IDP with the IDP matrix $G_{\text{net}}^{'}$. The symmetric part $\bar{G}_{\text{net}}^{'}$ of the IDP matrix can be used to estimate the system-wide stability by leveraging Theorems 1 and 2.

### 4.3 Two-level distributed stability assessment

The ODPMs and IDPMs derived above are exact differential passivity matrices in the sense of Definitions 2 and 3, and are not required to be symmetric in general. However, the stability criteria in Theorems 1 and 2 are established under symmetric matrix conditions. Therefore, in the practically relevant special case where no asymmetric coupling terms are present on the device side and the network is lossless, the exact ODPMs and IDPMs are symmetric, and the proposed stability criteria are strictly applicable. In more general engineering settings, such as those involving device-side cross-loop control or network losses, the exact differential passivity matrices are generally nonsymmetric. In these cases, their symmetric parts are used when applying Theorems 1 and 2 for approximate stability assessment.

**Remark 1**: Such a structured approximation is not uncommon in power-system analysis. A representative example arises in the construction of Lyapunov or energy functions for lossy power systems, where conductance-related terms generally make the associated vector field non-conservative and thus hinder the derivation of a globally well-defined energy function [31]. To address this difficulty, engineering-oriented approximations such as path-dependent assumptions and numerical energy functions have been widely adopted in practice [32]–[34]. These approximations share the same methodological spirit with the symmetrization used in this paper: when the exact analytical object does not possess the structure required by a desired stability-analysis tool, a structured approximation is introduced to recover a tractable and practically useful assessment framework.

Based on the above analysis, the semi-distributed stability criterion derived from Theorem 1 allows individual subsystems to compute their differential passivity matrices independently in a distributed manner. The control center then centrally verifies the system-wide stability based on the collected indices. Here, a two-level distributed stability assessment procedure is presented below, as shown in Fig. 5(A).

1) Device level: Obtain the linearized model of each device around the equilibrium, and determine the corresponding differential passivity matrix $\mathcal{S}_{\text{dev},i}$. The resulting matrices are then reported to the control center.

2) System level: Based on the network power flow, the control center constructs the IDP matrix $\bar{\mathcal{S}}_{\text{net}} = \bar{G}_{\text{net}}$ and evaluates the overall stability by verifying the sufficient condition $\bar{\mathcal{S}}_{\text{net}} + \bar{\mathcal{S}}_{\text{dev}} \succ 0$.

This passivity-based approach features streamlined data communication and reduced computational burden. Although eigenvalue computation is still required when verifying the condition $\overline{\mathcal{S}}_{\mathrm{net}} + \overline{\mathcal{S}}_{\mathrm{dev}} \succ 0$, the matrix involved is of much lower dimension compared to the method based on eigenanalysis, since the control center does not require detailed dynamic models but only the passivity indices of individual devices. An additional benefit is that the proposed approach can inherently preserve the privacy of device-level data.

In addition, the process of the fully distributed assessment indicated by Theorem 2 is illustrated in Fig. 5(B). In this process, the control center only needs to broadcast a scalar passivity index of the network side, while each device verifies its local stability condition in a distributed manner, without collecting matrix-valued indices and verifying the centralized stability condition. Notably, the scalar-index-based criterion in [14] is a special case of the proposed fully distributed criterion when all local passivity matrices are diagonal.

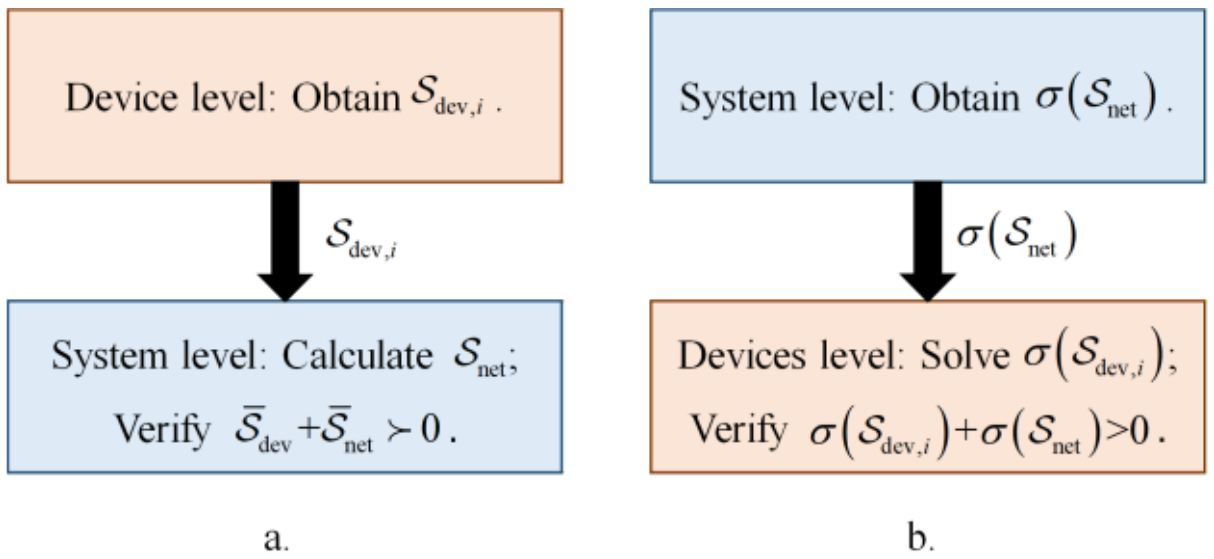

**Fig. 5 The application process of (a)** the semi-distributed stability criterion, **(b)** the fully distributed stability criterion.

## 5 Case study

### 5.1 Three-bus illustrative power system

As depicted in Fig. 6, a small-scale three-bus system, including one SG bus, one CD bus, and one QD bus, is used to validate the proposed criteria and methods. The device parameters of the three-bus system are provided in Table 1. The ODP matrices are analytically computed based on Propositions 1-3.

**Table 1** Device parameters of the three-bus system

| *Devices* | *Parameters* |
|---|---|
| SG | $M=0.1607, D=0.8, T_d=6.56, K_I=2.68,$ $x_d=0.295, x_{d'}=0.17, K_{\omega V}=0.066, K_{V\theta}=0.014,$ $K_{V\omega}=1.56$ |
| CD | $\tau_1=1, \tau_2=10, D_1=0.37, D_2=0.39,$ $K_{\omega V}=0.076, K_{V\theta}=0.076$ |
| QD | $\tau_1=0.8, \tau_2=8, D_1=0.37, D_2=0.37,$ $K_{\omega V}=0.078, K_{V\theta}=0.078$ |

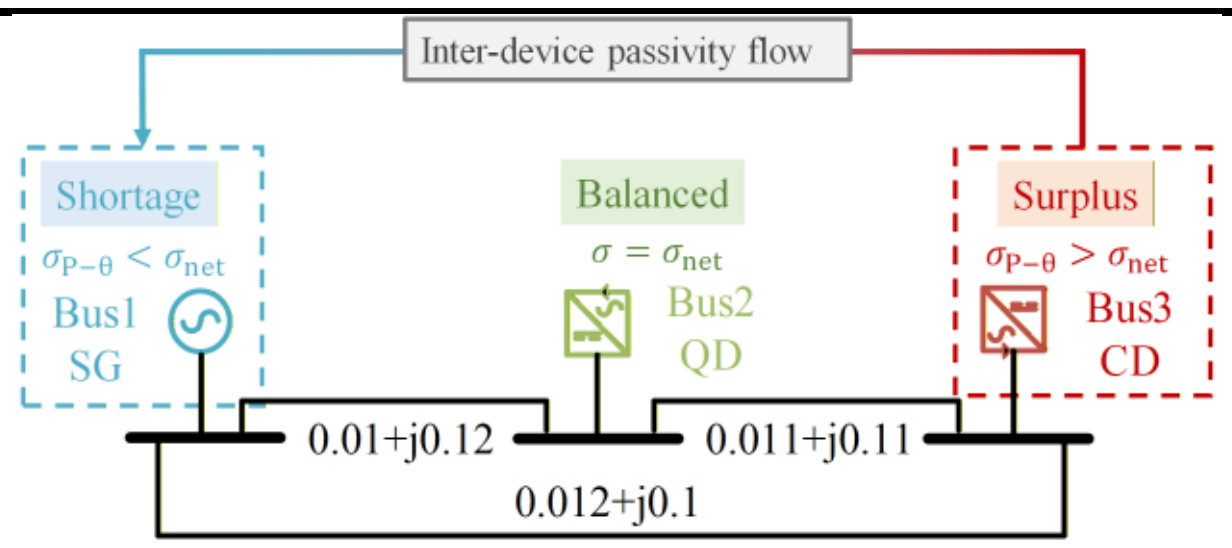

**Fig. 6 Topology of the three-bus system and passivity complementarity without cross-loop control.**

We first consider the case without cross-loop control. In the absence of cross-coupling, the ODP matrices become diagonal, resulting in decoupled $P$-$\theta$ and $Q$-$V$ channels. Let $\sigma_{P\text{-}\theta}$ and $\sigma_{Q\text{-}V}$ denote the differential passivity indices for the $P$-$\theta$ and $Q$-$V$ channels, respectively. The self-loop control parameters of SG and CD are tuned to vary $\sigma_{P\text{-}\theta}$, while other passivity indices are fixed to match the passivity index of the network.

The proposed criteria, implemented using the symmetric part of the differential passivity matrix, are evaluated under lossy networks. The minimum eigenvalue of the system IDP matrix $\mathcal{S}_{\text{net}}$ is $\sigma(\mathcal{S}_{\text{net}}) = -2.6769$. The stability verification results are illustrated in Fig. 7. In each figure, the blue area indicates the unstable region obtained by eigen-analysis, while the red area indicates the stability region identified by semi-distributed criterion. The shaded red area represents the distributed stability region predicted by fully-distributed criterion. The close alignment between the predicted and actual stability boundaries highlights the high accuracy of the proposed semi-distributed criterion. Moreover, a comparison between the matrix-valued passivity index–based criterion (Theorem 1) and the scalar-valued counterpart (Theorem 2) shows that using matrix-valued passivity indices improves precision and reduces conservativeness in stability assessment.

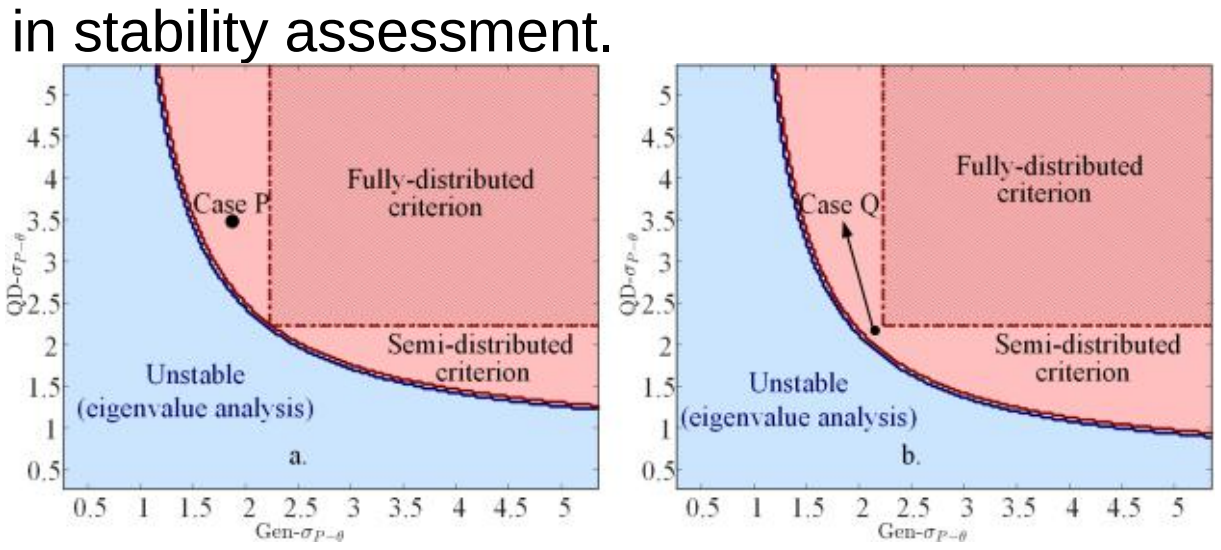

**Fig. 7 Small-signal stability regions of the three-bus system identified by eigen-analysis and the proposed criteria: (a)** without cross-loop control**; (b)** with cross-loop control**.**

To illustrate the complementarity mechanism, we examine the parameter setting marked as "Case P" in Fig. 7(a). The corresponding passivity matrices of three dynamical devices are shown below:

$$\mathcal{S}_1 = \begin{bmatrix} 1.86 & 0 \\ 0 & 2.68 \end{bmatrix}, \mathcal{S}_2 = \begin{bmatrix} 2.68 & 0 \\ 0 & 2.68 \end{bmatrix}, \mathcal{S}_3 = \begin{bmatrix} 3.47 & 0 \\ 0 & 2.68 \end{bmatrix}$$

As shown in Fig. 6, the $P-\theta$ channel of SG exhibits a passivity shortage, CD shows a surplus, and QD remains balanced. Although some device channels lack sufficient passivity, the system remains stable due to inter-device passivity compensation, which cannot be captured by scalar-valued passivity indices.

Next, the cross-loop control (18) and (22) are applied under lossy network conditions with fixed $K_{v\theta}$ and $K_{\omega v}$, and the results are shown in Fig. 7(b). It is observed that cross-loop control increases the conservativeness of the fully distributed criterion, whereas the semi-distributed criterion remains accurate. This stems from the fact that the fully distributed criterion utilizes the minimum eigenvalue of the passivity matrix, thereby neglecting the coupling between different input-output channels.

To further illustrate the coupling mechanism, we examine the parameter setting marked as "Case Q" in Fig. 7(b). The corresponding passivity matrices of three dynamical devices are shown below:

$$\mathcal{S}_1 = \begin{bmatrix} 2.15 & 0.07 \\ 0.11 & 2.68 \end{bmatrix}, \mathcal{S}_2 = \begin{bmatrix} 2.68 & 0.21 \\ 0.21 & 2.68 \end{bmatrix}, \mathcal{S}_3 = \begin{bmatrix} 2.15 & 0.16 \\ 0.22 & 2.68 \end{bmatrix}$$

Although diagonal entries of all device-level passivity matrices do not compensate for network side index $\sigma_{\text{net}} = -2.68$, indicating that the internal passivity of each individual $P-\theta$ and $Q$-$V$ channel is insufficient, the system remains stable. This is enabled by the presence of cross-channel coupling, which allows the system to satisfy passivity in the MIMO sense. It is worth noting that such internal coupling is not always beneficial to system stability, highlighting the need for appropriate control design and evaluation. The proposed framework provides a solid basis for this purpose.

### 5.2 IEEE-118 bus system with heterogeneous devices

We evaluate the proposed criteria in the modified IEEE 118-bus system, which comprises 41 SG buses, 18 CD buses, and 15 QD buses. The remaining buses are modeled as interconnection/load buses without dynamic device models. The basic network and power flow settings come from [35]. To examine how system passivity varies under changing operating conditions, the total load is scaled by a load scenario index $s$. In each scenario, the diagonal elements of the ODP matrices (as defined in Propositions 1–3) are set to the minimum values that satisfy the semi-distributed stability criterion. The cross-control parameters remain unchanged.

Fig. 8(a) compares the stability boundaries computed based on the proposed passivity-based semi-distributed criterion (Theorem 1) with those obtained by eigen-analysis. The y-axis shows the minimum eigenvalue of the ODP matrix $S_{\text{dev}}$ at marginal stability, obtained via either the passivity-based or eigenvalue-based method. It is observed that the two methods yield closely aligned stability boundaries. The discrepancy increases slightly at higher load levels, indicating a slightly larger conservativeness of the passivity-based criterion under heavier load. Such mild conservatism is acceptable in engineering practice, as power system operation prioritizes security over tight stability margins.

Similarly to Fig. 7, we further compare the stability results of the proposed criteria with those obtained by eigen-analysis under varying control parameters at $s=1$. Specifically, the self-control parameters of the devices are adjusted to represent different parameter scenarios. The comparison of stability regions, shown in Fig. 8(b), demonstrates the high accuracy and low conservativeness of the proposed criterion.

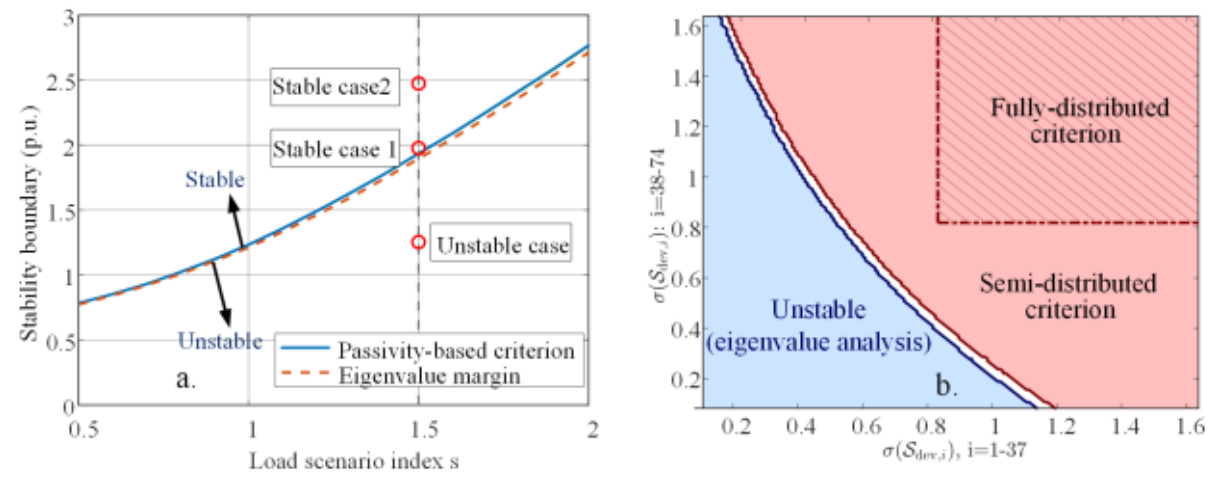

**Fig. 8 Comparison of stability boundaries under different load levels (a) and stability regions of the IEEE 118-bus system identified by eigenanalysis and the proposed criterion (b).**

We further justify the proposed criteria and methods within the context of nonlinear dynamics. Here, a small ground fault is applied at bus 5 with a fault resistance of 0.5 p.u. The trajectories of $\Delta\theta$ and the corresponding minimum eigenvalue $\sigma_{\text{dev},i}$ of the passivity indices $\mathcal{S}_{\text{dev},i}$ for the three representative cases are depicted in Figs. 9. In the figures illustrating the passivity indices, the outer circle labels represent the device IDs, while the inner radial values correspond to the scalar passivity indices of each device. The dashed red circle indicates the IDP index of the network. The transient responses suggest that larger values of the device-level passivity indices are generally associated with improved dynamic performance and stronger stability support in these cases.

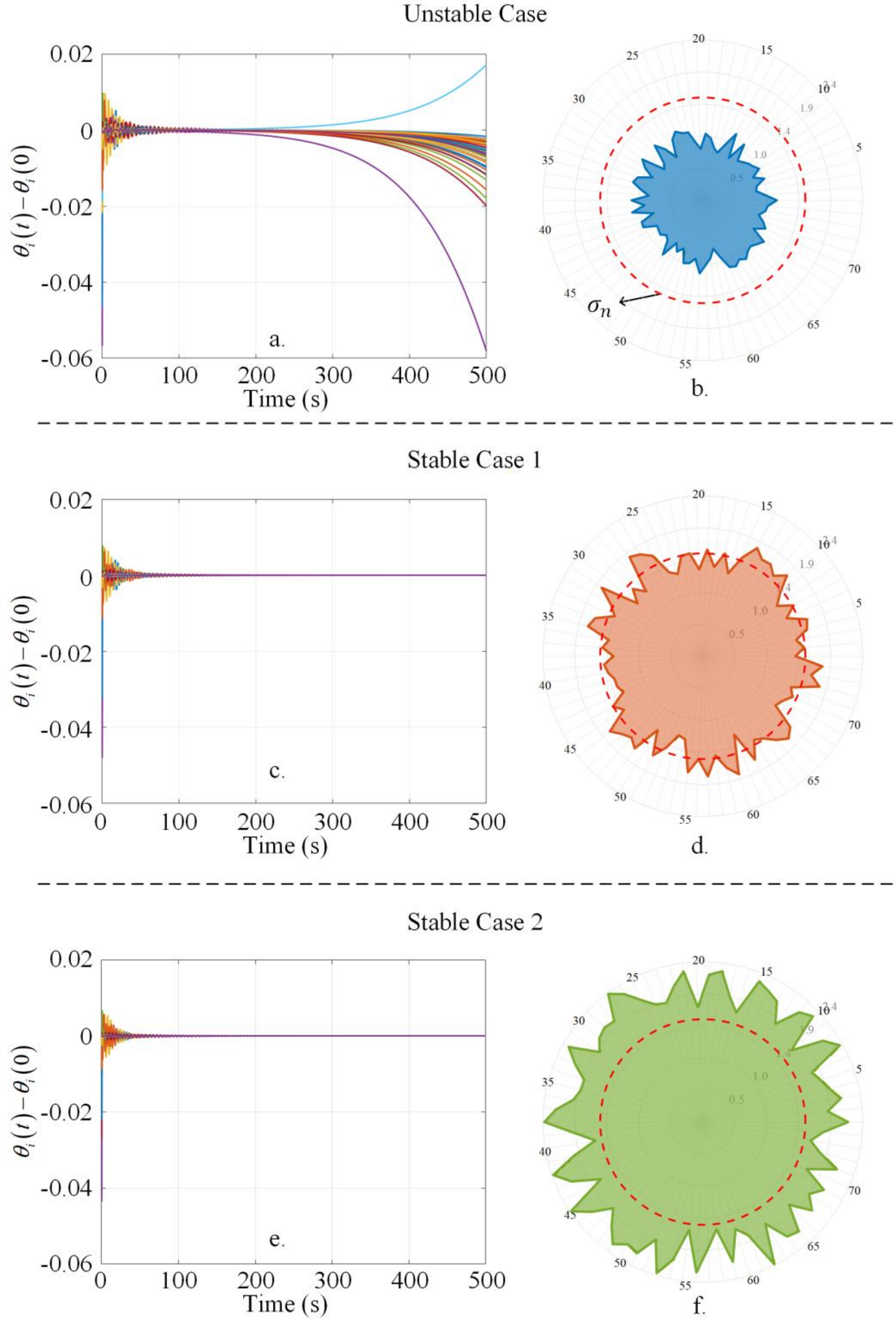

**Fig. 9 Simulation results of three cases with varying passivity indices. (a), (c) and (e) show the trajectories of phase deviations under the Unstable Case and Stable Cases 1-2, respectively. (b), (d) and (f) give the distributions of the minimum eigenvalues of the matrix-valued passivity indices.**

## 6 Conclusion

This paper presents a **matrix-valued differential passivity index** and the corresponding distributed stability analysis framework. As an extension of conventional scalar indices, the proposed matrix-valued index characterizes passivity in MIMO systems more comprehensively by capturing inter-channel coupling and passivity complementarity. Based on this formulation, system-level stability criteria are established and further decomposed into device-level conditions. The criteria rely on the passivity complementarity principle: system stability is ensured when the passivity surplus of devices compensates for the passivity shortage of the network. Compared with conventional scalar-index-based methods, the matrix-valued approach yields higher accuracy and lower conservatism in stability assessment.

In practical implementation, the stability assessment is carried out using the symmetric parts of the computed differential passivity matrices as an approximate treatment. The proposed matrix-valued differential-passivity-based stability assessment provides a theoretical basis for operating-point-dependent stability evaluation and stability identification in large-scale power systems, and its broader applications will be further explored in future work.

## Appendix

The Jacobian matrix blocks in (27) are shown below:

$$A = \frac{\partial P}{\partial \theta} = [A_{ij}] = \begin{cases} \sum_{k \neq i} V_i V_k B_{ik} \cos\theta_{ik}, i = j \\ -V_i V_j B_{ij} \cos\theta_{ij}, i \neq j \end{cases}$$

$$D = \frac{\partial P}{\partial V} = [D_{ij}] = \begin{cases} \sum_{k \neq i} V_k B_{ik} \sin\theta_{ik}, i = j \\ V_i B_{ij} \sin\theta_{ij}, i \neq j \end{cases}$$

$$C = \frac{\partial Q/V}{\partial V} = [C_{ij}] = \begin{cases} -B_{ij}, i = j \\ -B_{ij} \cos\theta_{ij}, i \neq j \end{cases}$$

The Jacobian matrix blocks in (28) are shown below:

$$
\begin{aligned}
A' &= \frac{\partial P}{\partial \theta} = [A'_{ij}] \\
&= \begin{cases} \sum_{k \neq i} V_i V_k (B_{ik} \cos\theta_{ik} - G_{ik} \sin\theta_{ik}), i = j \\ V_i V_j (-B_{ij} \cos\theta_{ij} + G_{ij} \sin\theta_{ij}), i \neq j \end{cases} \\
D' &= \frac{\partial P}{\partial V} = [D'_{ij}] \\
&= \begin{cases} 2G_{ii} V_i + \sum_{k \neq i} V_k (B_{ik} \sin\theta_{ik} + G_{ik} \cos\theta_{ik}), i = j \\ V_i (B_{ij} \sin\theta_{ij} + G_{ij} \cos\theta_{ij}), i \neq j \end{cases} \\
E' &= \frac{\partial (Q/V)}{\partial \theta} = [E'_{ij}] \\
&= \begin{cases} \sum_{k \neq i} V_k (B_{ik} \sin\theta_{ik} + G_{ik} \cos\theta_{ik}), i = j \\ -V_j (B_{ij} \sin\theta_{ij} + G_{ij} \cos\theta_{ij}), i \neq j \end{cases} \\
C' &= \frac{\partial Q/V}{\partial V} = [C'_{ij}] \\
&= \begin{cases} -B_{ij}, i = j \\ -B_{ij} \cos\theta_{ij} + G_{ij} \sin\theta_{ij}, i \neq j \end{cases}
\end{aligned}
$$

**Acknowledgements**

This research is supported by the Science and Technology Project of China Southern Power Grid Co., Ltd under Grant 036000KC23090004 (GDKJXM20231026).

**Declaration of competing interest**

The authors declare that they have no known competing financial interests or personal relationships that could have appeared to influence the work reported in this paper.

**References**

[1] F. Milano, *Power system modelling and scripting*. Springer Science & Business Media, 2010.
[2] M. J. Gibbard, P. Pourbeik, and D. J. Vowles, *Small-signal stability, control and dynamic performance of power systems*. University of Adelaide press, 2015.
[3] A. R. Bergen and D. J. Hill, “A Structure Preserving Model for Power System Stability Analysis,” *IEEE Transactions on Power Apparatus and Systems*, vol. PAS-100, no. 1, pp. 25–35, Jan. 1981, doi: 10.1109/TPAS.1981.316883.
[4] P. Kundur, “Power system stability,” *Power system stability and control*, vol. 10, no. 1, pp. 7–1, 2007.
[5] L. Meegahapola, A. Sguarezi, J. S. Bryant, M. Gu, E. R. Conde D., and R. B. A. Cunha, “Power System Stability with Power-Electronic Converter Interfaced Renewable Power Generation: Present Issues and Future Trends,” *Energies*, vol. 13, no. 13, p. 3441, Jul. 2020, doi: 10.3390/en13133441.
[6] M. A. Basit, S. Dilshad, R. Badar, and S. M. Sami ur Rehman, “Limitations, challenges, and solution approaches in grid-connected renewable energy systems,” *International Journal of Energy Research*, vol. 44, no. 6, pp. 4132–4162, 2020.
[7] S. Impram, S. V. Nese, and B. Oral, “Challenges of renewable energy penetration on power system flexibility: A survey,” *Energy strategy reviews*, vol. 31, p. 100539, 2020.
[8] J. Shair, H. Li, J. Hu, and X. Xie, “Power system stability issues, classifications and research

prospects in the context of high-penetration of renewables and power electronics," *Renewable and Sustainable Energy Reviews*, vol. 145, p. 111111, Jul. 2021, doi: 10.1016/j.rser.2021.111111.
[9] H. K. Khalil, *Nonlinear systems*. Prentice Hall, 2002.
[10] R. Estrada and C. Desoer, "Passivity and stability of systems with a state representation," *International Journal of Control*, vol. 13, no. 1, pp. 1–26, 1971.
[11] P. Borja, J. Ferguson, and A. Van Der Schaft, "Interconnection Schemes in Modeling and Control," *IEEE Control Syst. Lett.*, vol. 7, pp. 2287–2292, 2023, doi: 10.1109/LCSYS.2023.3286124.
[12] C. Spanias, P. Aristidou, and M. Michaelides, "A Passivity-Based Framework for Stability Analysis and Control Including Power Network Dynamics," *IEEE Systems Journal*, vol. 15, no. 4, pp. 5000–5010, Dec. 2021, doi: 10.1109/JSYST.2020.3007582.
[13] X. He and F. Dörfler, "Passivity and Decentralized Stability Conditions for Grid-Forming Converters," *IEEE Trans. Power Syst.*, pp. 1–4, 2024, doi: 10.1109/TPWRS.2024.3360707.
[14] P. Yang, F. Liu, Z. Wang, and C. Shen, "Distributed Stability Conditions for Power Systems With Heterogeneous Nonlinear Bus Dynamics," *IEEE Transactions on Power Systems*, vol. 35, no. 3, pp. 2313–2324, May 2020, doi: 10.1109/TPWRS.2019.2951202.
[15] F. Zhu, M. Xia, and P. J. Antsaklis, "Passivity analysis and passivation of feedback systems using passivity indices," in *2014 American Control Conference*, Jun. 2014, pp. 1833–1838. doi: 10.1109/ACC.2014.6858850.
[16] M. Li, L. Su, and G. Chesi, "Consensus of Heterogeneous Multi-Agent Systems With Diffusive Couplings via Passivity Indices," *IEEE Control Systems Letters*, vol. 3, no. 2, pp. 434–439, Apr. 2019, doi: 10.1109/LCSYS.2019.2893490.
[17] M. Xia, P. J. Antsaklis, and V. Gupta, "Passivity indices and passivation of systems with application to systems with input/output delay," in *53rd IEEE Conference on Decision and Control*, Los Angeles, CA, USA: IEEE, Dec. 2014, pp. 783–788. doi: 10.1109/CDC.2014.7039477.
[18] N. Kottenstette, M. J. McCourt, M. Xia, V. Gupta, and P. J. Antsaklis, "On relationships among passivity, positive realness, and dissipativity in linear systems," *Automatica*, vol. 50, no. 4, pp. 1003–1016, Apr. 2014, doi: 10.1016/j.automatica.2014.02.013.
[19] L. Harnefors, X. Wang, A. G. Yepes, and F. Blaabjerg, "Passivity-Based Stability Assessment of Grid-Connected VSCs—An Overview," *IEEE Journal of Emerging and Selected Topics in Power Electronics*, vol. 4, no. 1, pp. 116–125, Mar. 2016, doi: 10.1109/JESTPE.2015.2490549.
[20] S. Plesnick and P. Singh, "A Generalized Passivity-Based Stability Criterion for Assessing Large Signal Stability of Interconnected DC Power Systems," *IEEE Transactions on Aerospace and Electronic Systems*, vol. 59, no. 1, pp. 30–38, Feb. 2023, doi: 10.1109/TAES.2022.3189328.
[21] G. Wu *et al.*, "Passivity-Based Stability Analysis and Generic Controller Design for Grid-Forming Inverter," *IEEE Transactions on Power Electronics*, vol. 38, no. 5, pp. 5832–5843, May 2023, doi: 10.1109/TPEL.2023.3237608.
[22] X. Ru, P. Yang, F. Liu, and H. Mao, "Distributed stability analysis for power systems under persistent disturbance," in *2022 41st Chinese Control Conference (CCC)*, IEEE, 2022, pp. 6160–6165.
[23] K. Dey and A. M. Kulkarni, "Passivity-Based Decentralized Criteria for Small-Signal Stability of Power Systems with Converter-Interfaced Generation," *IEEE Transactions on Power Systems*, pp. 1–14, 2022, doi: 10.1109/TPWRS.2022.3192302.
[24] X. Ru, X. Peng, X. Chen, Z. Wang, P. Yang, and F. Liu, "Matrix-Valued Passivity Indices: Foundations, Properties, and Stability Implications," *arXiv preprint arXiv:2601.04796*, 2026.
[25] T. Stegink, C. De Persis, and A. van der Schaft, "A unifying energy-based approach to stability of power grids with market dynamics," *IEEE Transactions on Automatic Control*, vol. 62, no. 6, pp. 2612–2622, 2016.
[26] Y. Zhang and L. Xie, "Online Dynamic Security Assessment of Microgrid Interconnections in Smart Distribution Systems," *IEEE Transactions on Power Systems*, vol. 30, no. 6, pp. 3246–3254, Nov. 2015, doi: 10.1109/TPWRS.2014.2374876.
[27] J. W. Simpson-Porco, F. Dörfler, and F. Bullo, "Voltage Stabilization in Microgrids via Quadratic Droop Control," *IEEE Transactions on Automatic Control*, vol. 62, no. 3, pp. 1239–1253, Mar. 2017, doi: 10.1109/TAC.2016.2585094.
[28] M. Farrokhabadi, C. A. Cañizares, and K. Bhattacharya, "Frequency control in isolated/islanded microgrids through voltage regulation," *IEEE Transactions on Smart Grid*, vol. 8, no. 3, pp. 1185–

1194, 2015.

[29] W. Zhong, G. Tzounas, and F. Milano, “Improving the power system dynamic response through a combined voltage-frequency control of distributed energy resources,” *IEEE Transactions on Power Systems*, vol. 37, no. 6, pp. 4375–4384, 2022.

[30] X. Peng, C. Fu, P. Yang, X. Ru, and F. Liu, “Impact of Angle-voltage Coupling on Stability of Inverter-dominated Power Systems: A Damping Perspective,” *Authorea Preprints*, 2024.

[31] H.-D. Chiang, “Study of the existence of energy functions for power systems with losses,” *IEEE Transactions on Circuits and Systems*, vol. 36, no. 11, pp. 1423–1429, Nov. 1989, doi: 10.1109/31.41298.

[32] H.-D. Chang, C.-C. Chu, and G. Cauley, “Direct stability analysis of electric power systems using energy functions: theory, applications, and perspective,” *Proceedings of the IEEE*, vol. 83, no. 11, pp. 1497–1529, 1995, doi: 10.1109/5.481632.

[33] Y.-H. Moon, B.-H. Cho, T.-H. Rho, and B.-K. Choi, “The development of equivalent system technique for deriving an energy function reflecting transfer conductances,” *IEEE Transactions on Power Systems*, vol. 14, no. 4, pp. 1335–1341, Nov. 1999, doi: 10.1109/59.801893.

[34] H.-D. Chiang, J. Tong, and Y. Tada, “On-line transient stability screening of 14,000-bus models using TEPCO-BCU: Evaluations and methods,” in *IEEE PES General Meeting*, 2010, pp. 1–8. doi: 10.1109/PES.2010.5590026.

[35] R. D. Zimmerman, C. E. Murillo-Sánchez, and R. J. Thomas, “MATPOWER: Steady-state operations, planning, and analysis tools for power systems research and education,” *IEEE Transactions on Power Systems*, vol. 26, no. 1, pp. 12–19, 2010.